\documentclass[journal,10pt,onecolumn,draftclsnofoot]{IEEEtran}
\IEEEoverridecommandlockouts
\usepackage{amsmath,amssymb,amsfonts}
\usepackage{algorithmic}
\usepackage{algorithm}
\usepackage{array}
\usepackage[caption=false,font=normalsize,labelfont=sf,textfont=sf]{subfig}
\usepackage{textcomp}
\usepackage{stfloats}
\usepackage{url}
\usepackage{verbatim}
\usepackage{graphicx}
\usepackage{ntheorem}
\usepackage[algo2e]{algorithm2e} 
\usepackage{cite}
\usepackage{lipsum}
\usepackage{mathtools}
\usepackage{cuted}
\usepackage{setspace}
\usepackage{multicol}
\usepackage{epsfig}
\usepackage{epstopdf}
\usepackage{textcomp}
\usepackage{xcolor}
\usepackage{booktabs, multirow}
\usepackage{siunitx}

\usepackage{kantlipsum}
\usepackage{amssymb}
\usepackage{pifont}
\usepackage{tikz}
\usetikzlibrary{shapes, arrows, positioning}

\usepackage{float}
\usepackage{amsmath,amsfonts,amssymb}
\usepackage{bm}

\makeatletter

\newcommand{\Rmnum}[1]{\expandafter\@slowromancap\romannumeral #1@}
\def\BibTeX{{\rm B\kern-.05em{\sc i\kern-.025em b}\kern-.08em
    T\kern-.1667em\lower.7ex\hbox{E}\kern-.125emX}}

\begin{document}

\title{Pilot Assignment for Distributed Massive MIMO Based on Channel Estimation Error Minimization}

%\author{\IEEEauthorblockN{ Mohd Saif Ali Khan}
%\IEEEauthorblockA{\textit{SCEE, IIT Mandi} \\
%%\textit{Indian Institute of Technology,Mandi}\\
%Mandi, HP, India \\
%d21013@students.iitmandi.ac.in}
%\and
%\IEEEauthorblockN{ Karthik R.M.}
%\IEEEauthorblockA{\textit{ Ericsson India Pvt. Ltd.}\\
%Chennai, TN, India \\
%karthik.r.m@ericsson.com}
%\and
%\IEEEauthorblockN{ Samar Agnihotri}
%\IEEEauthorblockA{\textit{SCEE, IIT Mandi} \\
%%\textit{Indian Institute of Technology,Mandi}\\
%Mandi, HP, India \\
%samar@iitmandi.ac.in}
%}

\author{\IEEEauthorblockN{Mohd Saif Ali Khan* and Karthik R.M.$^+$, Samar Agnihotri*}\\
\IEEEauthorblockA{*School of Computing \& EE, Indian Institute of Technology Mandi, HP, India}\\
\IEEEauthorblockA{$^+$Ericsson India Pvt. Ltd., Chennai, TN, India}\\
Email: saifalikhan00100@gmail.com, r.m.karthik@gmail.com, samar.agnihotri@gmail.com
%
%\vspace{-0.3in}
}

        % <-this % stops a space
%\thanks{This paper was produced by the IEEE Publication Technology Group. They are in Piscataway, NJ.}% <-this % stops a space
%\thanks{Manuscript received April 19, 2021; revised August 16, 2021.}}



\maketitle

\begin{abstract}
Pilot contamination remains a major bottleneck in realizing the full potential of distributed massive MIMO systems. We propose two dynamic and scalable pilot assignment schemes designed for practical deployment in such networks. First, we present a low-complexity centralized scheme that sequentially assigns pilots to user equipments (UEs) to minimize the global channel estimation errors across serving access points (APs). This improves the channel estimation quality and reduces interference among UEs, enhancing the spectral efficiency. Second, we develop a fully distributed scheme that uses a priority-based pilot selection approach. In this scheme, each selected AP minimizes the channel estimation error using only local information and offers candidate pilots to the UEs. Every UE then selects a suitable pilot based on its AP priority. This approach ensures consistency and minimizes interference while significantly reducing pilot contamination. The method requires no global coordination, maintains low signaling overhead, and adapts dynamically to the UE deployment. Numerical simulations demonstrate the superiority of the proposed schemes in terms of network throughput when compared to the existing state-of-the-art schemes.
\end{abstract}

%\begin{IEEEkeywords}
%Distributed Massive MIMO,  Spectral Efficiency, Pilot Assignment, Distributed Algorithm
%\end{IEEEkeywords}

\section{Introduction}
\IEEEPARstart{D}{istributed} massive MIMO with cell-free architecture has emerged as a promising architecture for beyond-5G wireless networks, where a large number of geographically distributed APs coherently serve a smaller number of UEs using the same time-frequency resources \cite{ngo2017cell,khan2024joint}. This cell-free architecture eliminates cell boundaries and leverages favorable propagation and macro-diversity to improve coverage and spectral efficiency. However, its performance is fundamentally limited, among other factors, by pilot contamination, caused by the reuse of non-orthogonal pilots due to limited pilot resources within a coherence block.

Numerous pilot assignment (PA) strategies have been proposed to address pilot contamination. Earlier approaches such as random PA and greedy PA \cite{ngo2017cell}, and metaheuristic solutions like Tabu search-based PA \cite{liu2019tabu}, laid the foundation but are either suboptimal or computationally demanding. Structured techniques such as graph coloring PA (GC-PA) \cite{liu2020graph} improve spatial separation but lack adaptability in dynamic interference scenarios. Interference-aware PA (IA-PA)\cite{chen2022improving} improves fairness, and matching-based PA \cite{gao2023matching} enhances throughput over graph coloring and greedy baselines, though both suffer from high computational costs. The weighted graph framework PA (WGF-PA) \cite{zeng2021pilot} models pilot assignment as a Max $k$-Cut problem for global interference minimization but has quadratic complexity, making it impractical for large networks. Approximation-based PA (A-PA) \cite{wang2024pilot} achieves good performance but suffers from very high computational cost due to interior-point optimization, limiting its scalability. Approaches such as spectral clustering PA (SC-PA) \cite{zhang2023pilot} and anchor-based clustering PA (AC-PA)\cite{zhao2025fast} exploit graph partitioning and binary space partitioning, respectively. While they improve performance, they still incur high computational costs due to eigenvector calculation for the normalized graph Laplacians and $k$-means clustering for eigenvectors classification.

Most of these works are centralized and unsuitable for large and highly distributed networks. A few distributed pilot assignment schemes have been proposed such as survey propagation PA (SP-PA) \cite{kim2022survey}, which suffers from high complexity when the UE-to-pilot ratio is large, and distributed PA (D-PA) \cite{khan2024distributed} that offers a low-complexity parallel solution, but relies on iterative coordination, increasing the signaling overhead.  None of the well-performing PA schemes support dynamic environments (e.g., initial-access scenarios). In these methods, assigning a pilot to a newly arriving UE requires reassigning pilots to all previously connected UEs, which maybe infeasible for real-time operation. The scalable PA scheme \cite{bjornson2020scalable} offers low-complexity dynamic assignment, but it requires global transfer of each UE's PA to all APs, which may introduce significant signaling overhead and make the scheme infeasible in large-scale distributed systems.

A key research gap remains: A pilot assignment scheme that achieves a balance among low complexity, low overhead, scalability, and near-optimal performance in dynamic setting. In this paper, we propose two scalable and dynamic pilot assignment schemes for distributed massive MIMO (D-mMIMO) that address these demands. First, we propose a  \textbf{centralized scheme} that assigns pilots sequentially by minimizing  channel estimation error across each UE’s serving APs. It avoids revisiting prior assignments, enabling dynamic adaptability and low complexity. Then, a \textbf{distributed scheme} is proposed where each UE selects a set of priority APs in the descending order of large-scale fading coefficients (LSFCs). Each AP independently suggests pilot candidates by thresholding its local estimation error. Each UE finalizes the pilot via a priority-based rule, selecting a common pilot from the highest-priority AP proposing it. This approach reduces interference and ensures scalability through fully local decision-making.
We show that both schemes are scalable and significantly reduce pilot contamination in simulations, while maintaining practicality by supporting initial access, thus are suitable for large-scale deployments.

\textit{Organization}: The rest of the paper is organized as follows: Section II introduces the system model and corresponding closed-form spectral efficiency expression. Section III introduces the proposed centralised and distributed PA schemes. Section IV discusses performance through numerical simulations. Finally, Section V concludes the paper.

\section{System Model}
\label{system_model}
We consider a D-mMIMO network consisting of $M$ distributed APs, each with $A$ antennas, and $T$ single-antenna UEs. All APs jointly serve the UEs over the same time and frequency resources. A central processing unit (CPU) coordinates the APs via fronthaul connections, enabling joint signal processing. Channel estimation is performed using uplink pilots within a block-fading channel model, where each channel remains constant over a coherence block of $L_c$ symbols. Of these, $L_p$ symbols are reserved for pilot transmission. The wireless channel between AP $m$ and UE $t$ is modeled as $\mathbf{g}_{mt} {=} \sqrt{\beta_{mt}}\,\mathbf{h}_{mt} {\in} \mathbb{C}^{A \times 1}$, where $\beta_{mt}$ is the LSFC, and $\mathbf{h}_{mt} {\sim} \mathcal{CN}(\mathbf{0}, \mathbf{I}_A)$ models small-scale fading with Rayleigh distribution, assumed to be independent across APs and UEs. Let $\mathcal{T}_m$ denote the set of UEs served by AP $m$, and $\mathcal{M}_t$ denote the set of APs serving UE $t$. 

\subsection{Channel Estimation}
\label{channel_estimation}
During the uplink training phase, UE $t$ transmits a pilot sequence $\sqrt{L_p}\boldsymbol{\psi}_{i_t} {\in} \mathbb{C}^{L_p \times 1}$, where $i_t$ is the pilot index assigned to it, and $\|\boldsymbol{\psi}_{i_t}\|^2 {=} 1$. The pilot set consists of $L_p$ orthonormal sequences and multiple UEs may share the same pilot index when $T {>} L_p$, leading to pilot contamination. The pilot signal at AP $m$ is given by
\begin{align*}
    \mathbf{Y}^{\text{pilot}}_m = \sum_{t=1}^{T} \sqrt{p^p_t L_p} \mathbf{g}_{mt} \boldsymbol{\psi}_{i_t}^H + \mathbf{N}_m \in \mathbb{C}^{A \times L_p},
\end{align*}
where $p^p_t$ is the normalized signal-to-noise ratio (SNR) pilot power of UE $t$, and $\mathbf{N}_m$ is an additive white Gaussian noise (AWGN) matrix with i.i.d. entries drawn from $\mathcal{CN}(0,1)$. The minimum mean square error of the channel vector $\mathbf{g}_{mt}$ is then obtained as \cite{ngo2017cell}:
\begin{align*}
    \hat{\mathbf{g}}_{mt} = \frac{\gamma_{mt}}{\sqrt{p^p_t L_p} \beta_{mt}} \, \mathbf{Y}^{\text{pilot}}_m \boldsymbol{\psi}_{i_t},
\end{align*}
where
\begin{align*}
    \gamma_{mt} = \frac{p^p_t L_p \beta_{mt}^2}{\sum_{k=1}^{T} p^p_k L_p \beta_{mk} |\boldsymbol{\psi}_{i_t}^H \boldsymbol{\psi}_{i_k}|^2 + 1}.
\end{align*}
The $\gamma_{mt}$ reveals how pilot reuse  degrades channel estimation when multiple UEs using the same pilot interfere. 

\subsection{Uplink Data Transmission}
\label{uplink_transmission}
Let $x_t$ denote the uplink unit power data symbol transmitted by UE $t$. The uplink normalized SNR is denoted by $p^u_t$ for UE $t$. The received uplink signal at AP $m$ is 
\begin{align*}
    \mathbf{y}^u_m =  \sum_{t=1}^{T} \mathbf{g}_{mt} \sqrt{p^{u}_t} x_t + \mathbf{n}_m,
\end{align*}
where $\mathbf{n}_m \sim \mathcal{CN}(\mathbf{0}, \mathbf{I}_A)$ is the AWGN vector at AP $m$. Each AP performs local data detection using a linear received combining vector $\mathbf{v}_{mt} {\in} \mathbb{C}^{A \times 1}$ for each UE $t$. The locally estimated signal for UE $t$ at AP $m$ is then computed as $\hat{x}_{mt} {=} \mathbf{v}_{mt}^H \mathbf{y}^u_m$. After local processing, the soft estimates $\hat{x}_{mt}$ are forwarded to the CPU, where global combining is performed through large-scale fading decoding (LSFD). The final estimate $\hat{x}_t$ of UE $t$ is computed as
\begin{align}
\label{eq_1}
    \hat{x}_t = \sum\nolimits_{m=1}^{M} a_{mt} \hat{x}_{mt} = \sum\nolimits_{m=1}^{M} a_{mt} \mathbf{v}_{mt}^H \mathbf{y}^u_m,
\end{align}
where $a_{mt}$ denotes the LSFD weight applied at the CPU to the channel estimate from AP $m$ for UE $t$. By definition, \( a_{mt} {=} 0 \) if AP $m$ does not serve UE $t$; otherwise, it is computed using the optimal LSFD derived in \cite{zhang2021local,khan2025comments}.

In this work, we consider the use of the Partial Full-Pilot Zero-Forcing (PFZF) combining scheme for uplink signal detection, where the combining vectors are constructed locally at the APs\cite{zhang2021local}. The PFZF scheme suppresses interference from a selected subset $\mathcal{S}_m$ of strong UEs at AP $m$, while maintaining high array gain for the remaining UEs. Following local combining and LSFD processing of local data estimates, the effective SINR for UE $t$ is defined by 
\begin{align}
\label{eq_2}
\text{SINR}^{\text{PFZF}}_t = \frac{ p^{u}_t\Big|\sum\limits_{m=1}^{M}a^*_{mt}\sqrt{(A-\delta_{mt}L_{\mathcal{S}_m})\gamma_{mt}} \Big|^2}{\!\!\!\!\!\!\sum\limits_{k \in \mathcal{P}_t \backslash \{t\}}\!\!\!\!\!p^{u}_k\Big|\!\sum\limits_{m=1}^{M}\!\!a^*_{mt}\sqrt{(A-\delta_{mt}L_{\mathcal{S}_m})\gamma_{mk}}\Big|^2 + \sum\limits_{k=1}^{T}\!p^{u}_k\!\!\sum\limits_{m=1}^{M}\!\!|a_{mt}|^2(\beta_{mk} - \delta_{mt}\delta_{mk}\gamma_{mk}) + \sum\limits_{m=1}^{M}|a_{mt}|^2}.
\end{align}
where $\delta_{mt} {=}1$, when $t {\in} \mathcal{S}_m$ and $\delta_{mt} {=} 0$, otherwise.  The ergodic achievable uplink spectral efficiency (SE) of UE $t$ is lower-bounded as~\cite{zhang2021local,khan2025comments}:
\begin{align}
\label{eq_4}
\text{SE}^{u}_t = \left(\frac{1 - \frac{L_p}{L_c}}{2} \right) \log_2 \left(1 + \text{SINR}^{\text{PFZF}}_t \right).
\end{align}

\section{Pilot Assignment}
\label{PA}
In this section, we propose two scalable PA strategies based on minimizing channel estimation error. The first one follows a centralized approach, while the second leverages a distributed framework with local coordination among APs.

\subsection{Centralized Pilot Assignment}
\label{Cen_PA}
Efficient pilot assignment is essential in D-mMIMO systems to combat pilot contamination and improve the quality of channel estimation. Poor pilot reuse decisions may lead to overlapping pilot sequences among users with strong mutual interference, degrading the accuracy of the channel estimates and ultimately reducing the achievable spectral efficiency. We propose a scalable centralized scheme for pilot assignment based on minimizing the channel estimation error introduced by pilot contamination. The estimation error for UE $t$ is:
\begin{align}
\label{eq_5}
\text{Er}^{(i_t)}_t =\!\!\! \sum_{m \in \mathcal{M}_t} \Bigg( 
\frac{ p^p_t L_p\beta_{mt}^2}{p^p_t L_p\beta_{mt} + 1} 
- \frac{ p^p_t L_p\beta_{mt}^2}{\sum\limits_{k \in \mathcal{P}_{i_t}} p^p_k L_p\beta_{mk} + 1}
\Bigg),
\end{align}
where $\mathcal{P}_{i_t}$ denotes the set of UEs (including UE $t$) sharing the same pilot as UE $t$. The first term  in  \eqref{eq_5} represents the mean square of the channel estimate in the absence of pilot contamination, while the second term accounts for the contamination from all UEs in the set $\mathcal{P}_{i_t}$. Unlike previous studies that typically analyze pilot contamination between pairs of UEs, our proposed metric captures the aggregate contamination effect from all co-pilot UEs, providing a more comprehensive measure.
By minimizing $\text{Er}_t$ for each user, we indirectly maximize the effective SINR as shown in \eqref{eq_2}, since more accurate channel estimates improve both the desired signal gain and the   interference suppression. Following Algorithm~\ref{alg:centralized_pa}, running on the CPU, pilots are assigned to UEs sequentially in a greedy manner. For each incoming UE, the algorithm explores all available pilots and selects the one that minimizes the channel estimation error, considering the assignments of previously processed UEs. This scheme directly mitigates pilot contamination at each step, enhancing channel estimation accuracy and improving SINR performance. Moreover, the pilot assignment is performed iteratively for each UE without revisiting or modifying previously assigned pilots, making the approach highly suitable for dynamic and practical network deployments.
\begin{algorithm}[]
\caption{Estimation error minimization PA (EEM-PA) }
\label{alg:centralized_pa}
\begin{algorithmic}[1]
%\REQUIRE $T$,  $L_p$, large-scale fading matrix $\boldsymbol{\beta} $,  $\{p^p_t\}_{t=1}^T$
\ENSURE Pilot indices $\{i_1, i_2, \dots, i_T\}$ for all UEs
\STATE Initialize PA list: $\mathcal{I} = \emptyset$
\FOR{$t = 1$ to $T$}
    \IF{$t \leq L_p$}
        \STATE Assign unique pilot: $i_t = t$
    \ELSE
        \STATE Initialize minimum estimation error: $\text{Er}^{\text{min}}_t = \infty$
        \FOR{$i = 1$ to $L_p$}
            \STATE Let $\mathcal{P}_i = \{k < t : i_k = i\} \cup \{t\}$ 
            \STATE Compute $\text{Er}_t^{(i)}$ using \eqref{eq_5}.
            \IF{$\text{Er}_t^{(i)} < \text{Er}^{\text{min}}_t$}
                \STATE Update: $\text{Er}^{\text{min}}_t = \text{Er}_t^{(i)}$, $i_t = i$
            \ENDIF
        \ENDFOR
    \ENDIF
    \STATE Append $i_t$ to assignment list: $\mathcal{I} \gets \mathcal{I} \cup \{i_t\}$
\ENDFOR
%\RETURN $\mathcal{I} = \{i_1, i_2, \dots, i_T\}$
\end{algorithmic}
\end{algorithm}

\textit{Complexity Analysis} :
For each UE, the algorithm evaluates the estimation error for all \( L_p \) pilot candidates. Each estimation error computation involves a sum over the serving APs \( \mathcal{M}_t \); and for each AP, a sum over the UEs already assigned the same pilot. Since these pilot-sharing user groups are maintained incrementally, the algorithm avoids recomputing pilot contamination terms from scratch. As a result, the per UE computational cost is approximately \( \mathcal{O}(|\mathcal{M}_t| L_p) \). Importantly, this cost is independent of the  \( T \) and \( M \), ensuring scalability of the proposed method.

\subsection{Distributed Pilot Assignment}
\label{Dist_PA}

To enhance scalability and minimize pilot contamination in a fully distributed manner, we propose a UE-driven distributed PA scheme where each UE participates in selecting its own pilot with assistance from a small subset of its strongest serving APs. When UE $t$ enters the network, among serving APs, the top \({S}\) serving APs with the strongest LSFCs are selected to recommend pilots for UE $t$. Each selected AP $m$, for every pilot sequence \(i\) computes the local estimation error as given by: 
\begin{align}
\label{eq_6}
\text{Er}^{(i_t)}_{mt} =  \Bigg( 
\frac{ p^p_t L_p\beta_{mt}^2}{p^p_t L_p \beta_{mt} + 1} 
- \frac{p^p_t L_p \beta_{mt}^2}{\sum\limits_{k \in \mathcal{P}^{m}_{i_t}} p^p_k L_p \beta_{mk} + 1}
\Bigg),
\end{align}
where $\mathcal{P}^{m}_{i_t}$ denotes the set of UEs (including UE $t$) sharing the same pilot as UE $t$ and served by AP $m$. Similar to \eqref{eq_5}, the local matric in \eqref{eq_6} sees the  impact of pilot contamination locally at each AP by its serving UEs only.
Rather than suggesting a single pilot, each AP constructs a candidate pilot set comprising of all pilots whose estimation error is within a  relative threshold ($\Delta$) of the best-performing pilot. The steps running on APs are given in Algorithm \ref{alg:distributed_sequential}. This approach ensures that multiple candidates with low local channel estimation error are available, increasing the chance of reaching consensus across APs. The candidate pilot sets are then sent to UE $t$ by top $S$ selected APs.
\begin{algorithm}[]
\caption{Distributed Priority-Based PA (DPB-PA) }
\label{alg:distributed_sequential}
\begin{algorithmic}[1]
%\REQUIRE Large-scale fading matrix \(\boldsymbol{\beta}\),  $\{p^p_t\}_{t=1}^T$,  \(L_p\), threshold \(\Delta\)
\ENSURE Selected pilot \(i_t\) for the UE \(t\)
\FOR{each new UE \( t \)}
    \STATE Select top \( S \) APs: the APs with the highest \(\beta_{mt}\) values among \(\mathcal{M}_t \)
    \FOR{each selected AP \( m \in S \)}
        \FOR{each pilot \( i = 1 \) to \( L_p \)}
            \STATE Compute estimation error \( \text{Er}_{mt}^{(i)} \) using Eqn.~\eqref{eq_6}
        \ENDFOR
        \STATE Find minimum estimation error \(\text{Er}_{mt}^{\text{min}}\)
        \STATE Create candidate set:
        $
        \mathcal{C}_m = \left\{ i : \text{Er}_{mt}^{(i)} \leq (1+\Delta) \, \text{Er}_{mt}^{\text{min}} \right\}
        $
        \STATE Send \(\mathcal{C}_m\) to the UE \( t \)
    \ENDFOR
    
    \STATE Call Algorithm \ref{alg:priority} to select pilot \(i_t\)

\ENDFOR

\end{algorithmic}
\end{algorithm}

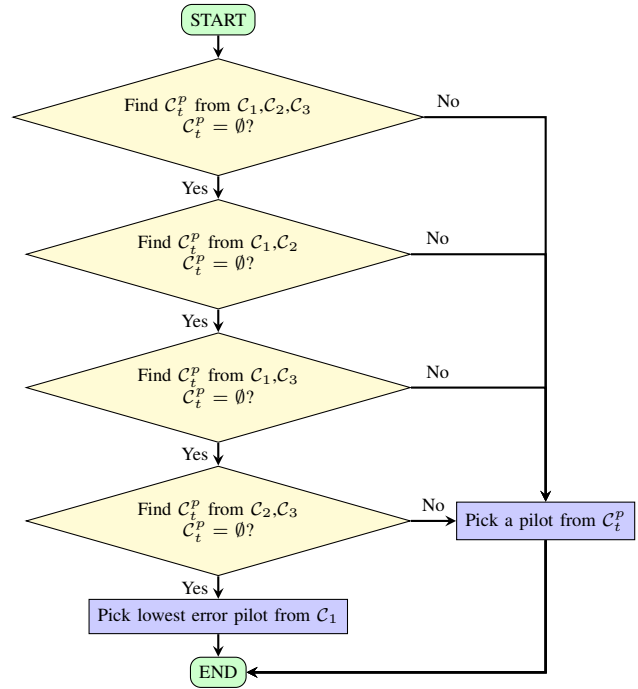
\begin{figure}[]
\centering
\begin{tikzpicture}[
    node distance=0.3cm and 0.6cm,
    startstop/.style={rectangle, rounded corners, minimum width=0.5cm, minimum height=0.3cm, text centered, draw=black, fill=green!20, font=\scriptsize},
    decision/.style={diamond, aspect=3.5, minimum width=0.8cm, minimum height=0.1cm, text centered, draw=black, fill=yellow!20, font=\scriptsize},
    process/.style={rectangle, minimum width=0.6cm, minimum height=0.3cm, text centered, draw=black, fill=blue!20, font=\scriptsize},
    arrow/.style={thick,->,>=stealth}
]

% Nodes
\node (start) [startstop] {START};
\node (check123) [decision, below=of start, align=center] {Find $\mathcal{C}^p_t$ from $\mathcal{C}_1$,$\mathcal{C}_2$,$\mathcal{C}_3$  \\ $\mathcal{C}^p_t = \emptyset$?};
\node (check12) [decision, below=of check123, align=center] {Find $\mathcal{C}^p_t$ from $\mathcal{C}_1$,$\mathcal{C}_2$ \\ $\mathcal{C}^p_t = \emptyset$?};
\node (check13) [decision, below=of check12, align=center] {Find $\mathcal{C}^p_t$ from $\mathcal{C}_1$,$\mathcal{C}_3$ \\ $\mathcal{C}^p_t = \emptyset$?};
\node (check23) [decision, below=of check13, align=center] {Find $\mathcal{C}^p_t$ from $\mathcal{C}_2$,$\mathcal{C}_3$ \\ $\mathcal{C}^p_t = \emptyset$?};
\node (pick1) [process, below=of check23] {Pick lowest error pilot from $\mathcal{C}_1$};
\node (end) [startstop, below=of pick1] {END};

% Processes
\node (pick123) [process, right= of check23] {Pick  a pilot from $\mathcal{C}^p_t$};
%\node (pick12) [process, right=of check12] {Pick  common pilot from $\mathcal{C}_1$};
%\node (pick13) [process, right=of check13] {Pick  common pilot from $\mathcal{C}_1$};
%\node (pick23) [process, right=of check23] {Pick  common pilot from $\mathcal{C}_2$};

% Arrows
\draw [arrow] (start) -- (check123);
\draw [arrow] (check123) -| node[pos=0.1, above, font=\scriptsize] {No} (pick123);
\draw [arrow] (pick123) |- (end);

\draw [arrow] (check123) -- node[left, font=\scriptsize] {Yes} (check12);
\draw [arrow] (check12) -| node[pos=0.1, above, font=\scriptsize] {No} (pick123);
\draw [arrow] (pick123) |- (end);

\draw [arrow] (check12) -- node[left, font=\scriptsize] {Yes} (check13);
\draw [arrow] (check13) -| node[pos=0.1, above, font=\scriptsize] {No} (pick123);
\draw [arrow] (pick123) |- (end);

\draw [arrow] (check13) -- node[left, font=\scriptsize] {Yes} (check23);
\draw [arrow] (check23) -- node[above, font=\scriptsize] {No} (pick123);
\draw [arrow] (pick123) |- (end);

\draw [arrow] (check23) -- node[left, font=\scriptsize] {Yes} (pick1);
\draw [arrow] (pick1) -- (end);

\end{tikzpicture}
\caption{Priority-Based Pilot Selection Flowchart}
\label{fig_pr}
\end{figure}

Upon receiving recommendations, UE $t$ finalizes its pilot assignment through a generalized priority-based selection mechanism. It first searches for common pilots ($\mathcal{C}^p_t$) among the candidate sets shared by all \(S\) APs. If one or more common pilots exist, it is assigned a pilot randomly from the common pilots. If no such pilot exists, it progressively relaxes the condition by seeking $\mathcal{C}^p_t$ among \(S{-}1\) APs, \(S{-}2\) APs, and so forth. This priority search continues until $\mathcal{C}^p_t$ becomes non-empty, ensuring both consistency across APs and minimization of local interference as seen from Fig. \ref{fig_pr} for $S{=}3$. If $\mathcal{C}^p_t$ remains empty, then the UE defaults to selecting the best pilot proposed by its highest-priority AP. The steps running on UEs are in Algorithm \ref{alg:priority}.

\begin{algorithm}[]
\caption{Priority-Based Pilot Selection at the UE}
\label{alg:priority}
\begin{algorithmic}[1]
%\REQUIRE Candidate pilot sets \(\mathcal{C}_1, \mathcal{C}_2, \ldots, \mathcal{C}_{\bar{M}}\)
\ENSURE Selected pilot \(i_t\) for the UE \(t\)

\FOR{level \( l = S \) down to \( 2 \)}
    \STATE Generate all $\binom{S}{l}$ AP groups of size $l$
   
    \FOR{each group}
        \STATE Find $\mathcal{C}^p_t$ among the selected APs
        \IF{$\mathcal{C}^p_t \neq \emptyset$}
            \STATE Select a pilot, \(i_t\), from $\mathcal{C}^p_t$ \COMMENT{Refer Fig. \ref{fig_pr}}
            \STATE \textbf{Break the procedure}
        \ENDIF
    \ENDFOR
 \ENDFOR 
%\STATE \textbf{Fallback:} Select a pilot from \(\mathcal{C}_1\) as \(i_t\)
%\STATE Assign selected pilot as \(i_t\)

\end{algorithmic}
\end{algorithm}
The estimation error is minimized for each UE with respect to its multiple serving APs, improving channel estimation quality and consequently the SINR. Although the performance does not match that of the centralized approach, since pilot assignment is based solely on local information, it is inherently dynamic: assigning a pilot to a new UE does not impact the pilots of already assigned UEs. The method effectively reduces pilot conflicts while distributing the selection workload, requiring only very low complexity operations at the UEs. By combining localized estimation error minimization with UE-driven final pilot selection through a priority-based rule, this scheme is well-suited for dynamic and large-scale D-mMIMO networks, as it proceeds sequentially in a fully distributed manner without any signaling overhead among APs or with the CPU, and without relying on centralized control.

\textit{Complexity Analysis} : For each UE, the estimation errors for all pilot candidates are computed, requiring \(\mathcal{O}(\bar{M} L_p)\) operations. At each UE, the common pilots are checked, leading to a complexity of \(\mathcal{O}(\bar{M}^2  L_p)\). Thus, the total complexity per UE is \(\mathcal{O}(\bar{M}  L_p + \bar{M}^2  L_p)\). Importantly, this per UE cost is independent of the  \( T \) and \( M \), ensuring scalability of the proposed method.

\subsection{Complexity and Overhead Comparison}
A comparison of the computation costs of different PA schemes is given in Table~\ref{tab:PA_complexity}. This table indicates that our proposed schemes have lesser complexity than their existing counterparts.
\begin{table}[]
\centering
\footnotesize
\caption{Computational Complexity of Different PA Schemes}
\begin{tabular}{|l|l|}
\hline
\textbf{PA Scheme} & \textbf{Complexity} \\
\hline
IA-PA \cite{chen2022improving} & $\mathcal{O}(T M^2 + M T^2)$ \\
GC-PA \cite{liu2020graph} & $\mathcal{O}(T M \log M + M T + T^2)$ \\
SC-PA \cite{zhang2023pilot} & $\mathcal{O}(|\mathcal{M}_t|T^2 + T^3)$ \\
AC-PA \cite{zhao2025fast} & $\mathcal{O}(|\mathcal{M}_t| T^2 + T^3)$ \\
A-PA \cite{wang2024pilot} & $\mathcal{O}(T^{3.5})$ \\
WGF-PA \cite{zeng2021pilot} & $\mathcal{O}(|\mathcal{M}_t| T^2)$ \\
\hline
\textbf{Proposed EEM-PA} & $\mathcal{O}(|\mathcal{M}_t| T L_p)$ \\
\hline
SP-PA \cite{kim2022survey} & $\mathcal{O}(e^{T})$ \\
Scalable PA \cite{bjornson2020scalable} & $\mathcal{O}(M T)$ \\
D-PA \cite{khan2024distributed} & $\mathcal{O}(M T + I L_p^2)$ \\
\hline
\textbf{Proposed DPB-PA} & $\mathcal{O}(T S L_p + T S^2 L_p)$ \\
\hline
\end{tabular}
\label{tab:PA_complexity}
\end{table}
For example, the near-optimal weighted graph framework PA (WGF-PA) \cite{zeng2021pilot}  has a complexity of \(\mathcal{O}(MT^2)\). In comparison, the proposed centralized estimation error minimization PA (EEM-PA) scheme has a significantly lower complexity of \(\mathcal{O}(|\mathcal{M}_t| T L_p)\), where both \(|\mathcal{M}_t|\) and \(L_p\) are very small compared to \(M\) and \(T\).

The SP-PA scheme~\cite{kim2022survey} suffers from high complexity and scalable-PA~\cite{bjornson2020scalable} suffers from high overhead since each AP must know the pilot assignments of all previously assigned UEs, leading to a heavy information exchange among APs. In D-PA \cite{khan2024distributed},  each AP shares pilot information with neighboring APs at each iteration, resulting in considerable overhead. In contrast, the proposed Distributed Priority-Based PA (DPB-PA)  requires no additional overhead, apart from UEs notifying their serving APs of the selected pilot, an essential step for local combining that is also necessary in the above-mentioned methods. Thus, the proposed DPB-PA scheme achieves low complexity without introducing extra signaling overhead.

\section{Numerical Simulations}
\label{simulations}
In our simulation setup \( M {=} 100 \) APs and \( T {=} 100 \) UEs are uniformly distributed within a square area of \(1 {\times} 1\)~km\(^2\). Each AP is equipped with \( A {=} 8 \) antennas. The channel bandwidth is set to \( B {=} 20 \)~MHz. The coherence block length is \( L_c {=} 200 \) symbols, with \( L_p {=} 7 \) symbols allocated for uplink pilot transmission, satisfying the requirement \( A {>} L_p \) for PFZF. We have considered $S{=}3$ and $\Delta{=}0.1$ (for best performance, the $\Delta$ should be near zero). The LSFCs are generated according to the model in \cite{ngo2017cell}, incorporating a shadow fading standard deviation of 8~dB. UEs are assigned APs using LSFCs in the same way as in \cite{liu2020graph}, with threshold $\beta_{th} {=} 0.95$. We assume that all UEs transmit with power of $100$~mW. Strong UE grouping is performed in the same way as in \cite{zhang2021local}.
\begin{figure}[]
\centering
\includegraphics[width=0.65\textwidth]{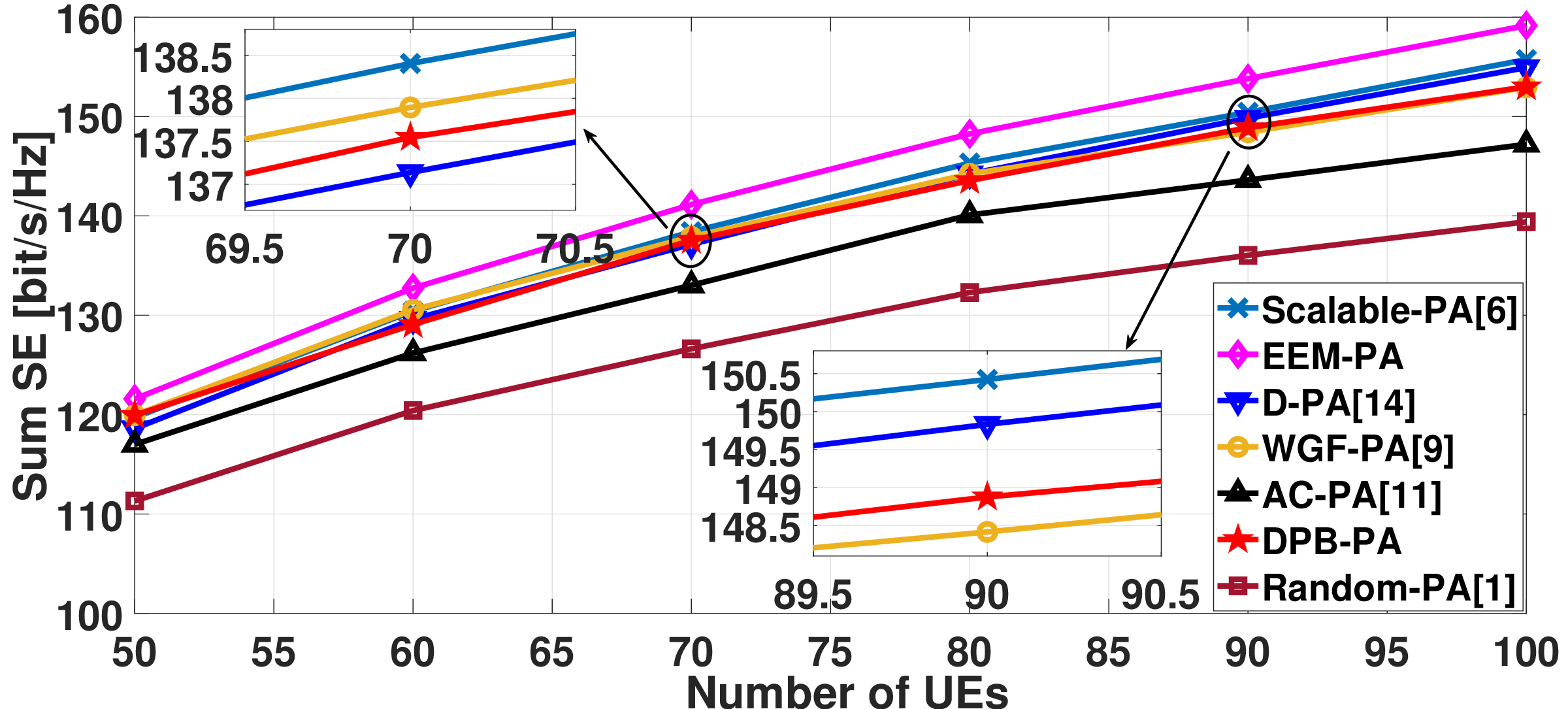}
\caption{The uplink sum SE vs number of UEs ($T$) plot for different PA schemes.}
\label{fig_1}
\end{figure}

Fig.~\ref{fig_1} shows the sum SE performance as the number of UEs increases from 50 to 100. As expected, the sum SE initially improves with increasing UEs due to enhanced spatial reuse, until inter-user interference begins to dominate. The proposed EEM-PA scheme consistently outperforms all benchmark schemes across the entire range of user densities, with the performance gap widening as the number of UEs increases.  This superior performance is attributed to the proposed channel estimation error metric in \eqref{eq_5}, which, unlike WGF-PA, AC-PA, SC-PA, and other conventional methods, considers pilot contamination jointly over all previously assigned UEs rather than only the pairwise interactions among the UEs. Although Scalable-PA also accounts for contamination from all previous UEs, it evaluates it only from the perspective of a single AP and relies on the sum of LSFCs among pilot-sharing UEs. In contrast, the EEM-PA uses a more precise estimation error-based metric, leading to superior pilot decisions.

Notably, the proposed DPB-PA scheme, despite being fully decentralized and operating solely on local information without inter-AP coordination, demonstrates robust and competitive performance compared to schemes such as WGF-PA, Scalable-PA, and D-PA. Both Scalable-PA and D-PA involve signaling overhead due to coordination among APs. The performance of the proposed DPB-PA scheme is due to local estimation error metric introduced in \eqref{eq_6}, which serves as a distributed counterpart to the centralized metric in \eqref{eq_5}. The DPB-PA scheme significantly outperforms the AC-PA and only trails behind the EEM-PA, as expected.
\begin{figure}[]
\centering
\includegraphics[width=0.65\textwidth]{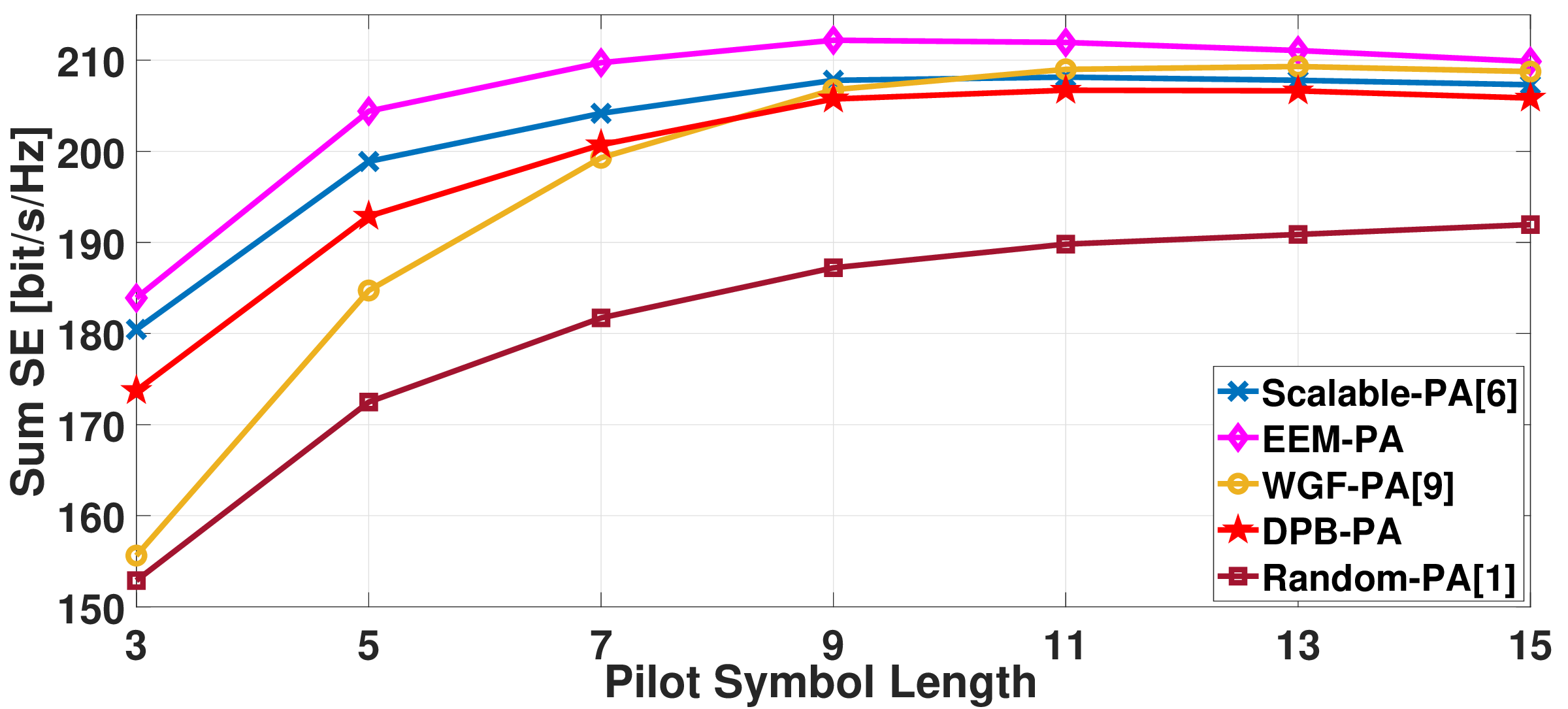}
\caption{The uplink sum SE vs pilot symbol length ($L_p$) plot for different PA schemes, when number of antennas per AP, $A=16$.}
\label{fig_2}
\end{figure}

Fig.~\ref{fig_2} illustrates the sum SE achieved by different PA strategies as pilot symbol length ($L_p$) increases. The sum SE  for all schemes first increases with increase in $L_p$,  but then degrades as fewer symbols remain for data transmission. The proposed EEM-PA consistently achieves the highest SE across all values of $L_p$. Similarly, the proposed DPB-PA demonstrates competitive performance with other benchmark schemes and remains relatively stable across different pilot lengths, unlike WGF-PA, which shows more sensitivity to shorter pilot lengths. 
\begin{figure}[]
\centering
\includegraphics[width=0.65\textwidth]{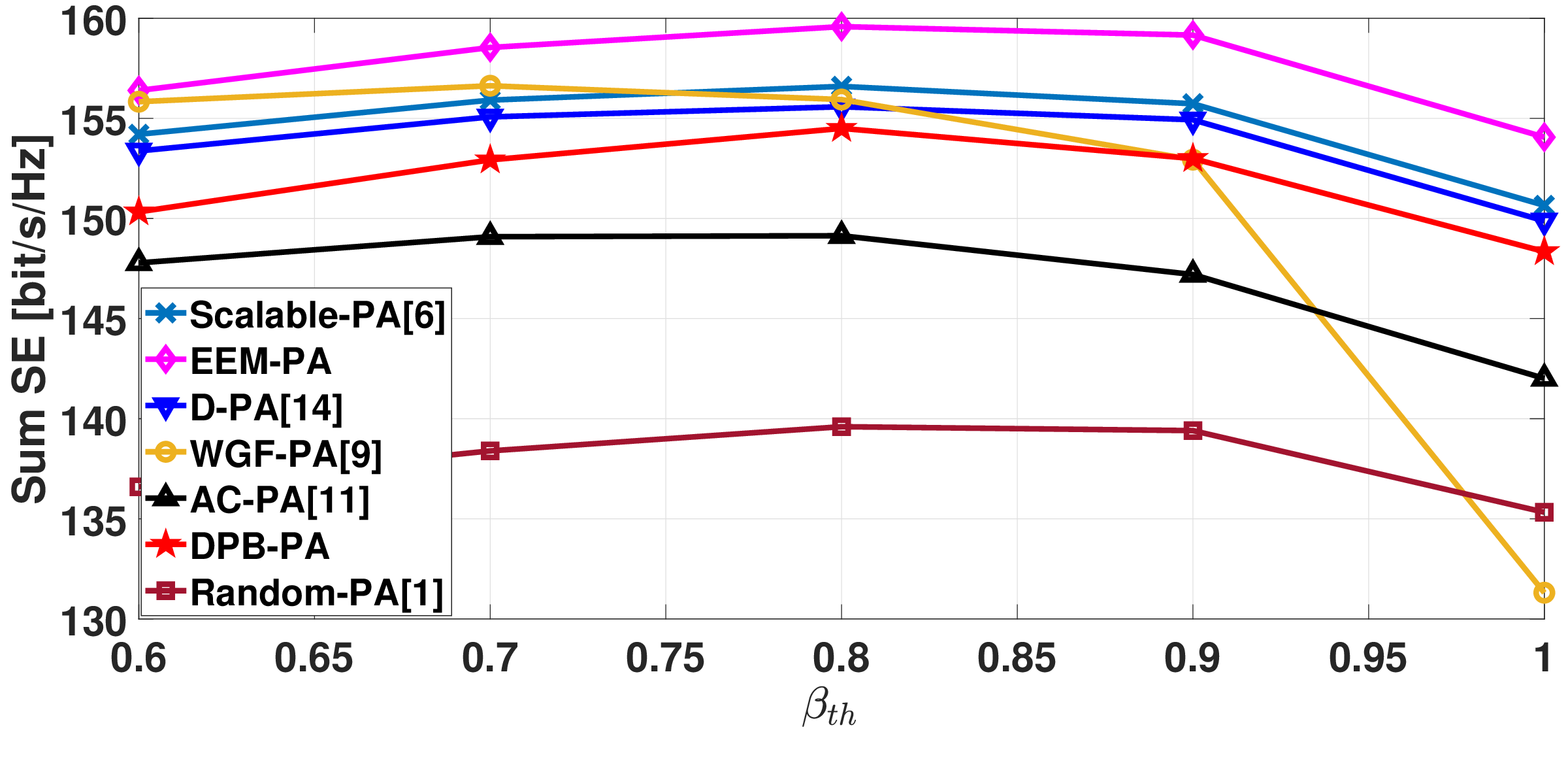}
\caption{The uplink sum SE vs AP-UE association threshold $\beta_{th}$.}
\label{fig_3}
\end{figure}

In D-mMIMO networks, a pilot assignment scheme must enhance the channel estimation quality between each UE and its associated APs. Therefore, it is critical to evaluate the PA performance under varying AP-UE association thresholds. Fig.~\ref{fig_3} illustrates the impact of the association threshold $\beta_{th}$ on sum SE.  As $\beta_{th}$ increases, more APs are associated with each UE, initially improving the sum SE due to increased spatial diversity. However, beyond a certain point, further increase leads to performance degradation owing to the inclusion of strongly interfering APs. The proposed EEM-PA consistently achieves the highest SE across all values of $\beta_{th}$. While the WGF-PA performs comparably at lower $\beta_{th}$, its performance significantly deteriorates at higher thresholds, especially as $\beta_{th} {\to} 1$. This is because the WGF-PA relies on spatial separation among serving APs to compute pilot contamination, which becomes ineffective when all APs are associated and spatial separation diminishes, leading to incorrect estimation of pilot contamination levels. In contrast, EEM-PA directly leverages a precise estimation error metric, remains robust under such conditions. Also, the proposed DPB-PA maintains strong performance across all values of  $\beta_{th}$. Its ability to effectively counteract pilot contamination remains largely unaffected by the increase in AP-UE associations.
\begin{figure}[]
\centering
\includegraphics[width=0.65\textwidth]{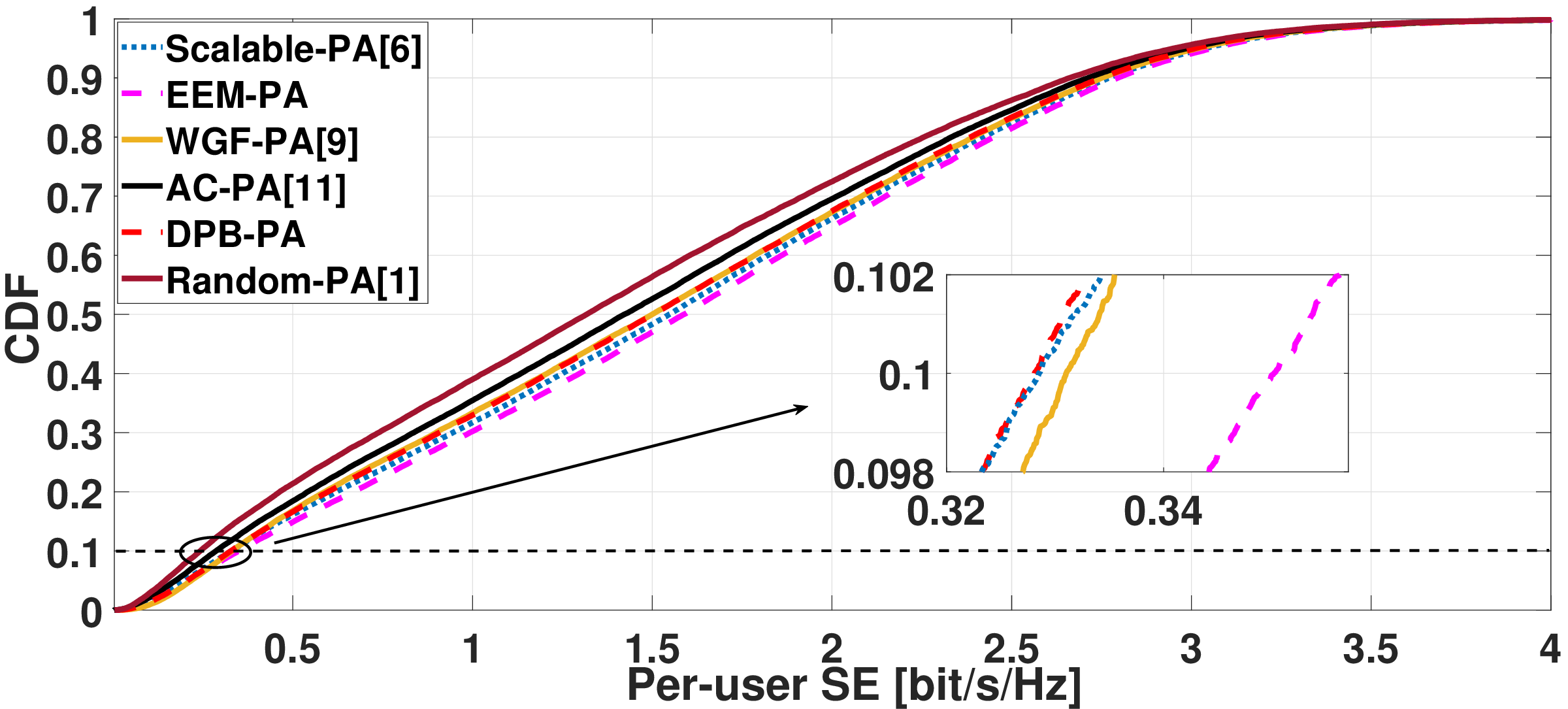}
\caption{Cumulative distribution function (CDF) versus Per-user SE for different PA schemes.}
\label{fig_4}
\end{figure}

Without any fairness criterion, improvements in sum SE can come at the expense of badly-performing UEs experiencing poor per-user SE. Therefore, a robust pilot assignment scheme must also ensure fairness by maintaining strong per-user SE performance. Fig.~\ref{fig_4} presents the CDF of per-user SE for different PA schemes. In terms of the $90\%$-likely SE (SE achieved by $90\%$ UEs), the proposed EEM-PA outperforms all existing strategies, indicating its effectiveness in enhancing not just the overall performance but also the SE of the weak UEs. The proposed DPB-PA also demonstrates comparable $90\%$-likely SE performance to WGF-PA and Scalable-PA, confirming its ability to balance sum SE and user fairness in a fully distributed setting.

\textit{Discussion:} The proposed EEM-PA and DPB-PA schemes demonstrate both high performance and strong robustness under diverse network configurations and system parameters. The EEM-PA scheme consistently achieves the highest sum SE across various scenarios, including varying UE densities, varying pilot lengths, and changing AP-UE association thresholds. Its superior performance is driven by a globally-aware estimation error metric that accounts for cumulative pilot contamination across all previously assigned UEs, enabling more accurate pilot selection. The DPB-PA scheme, despite being decentralized and relying only on local information without any inter-AP coordination, also achieves competitive SE. This is due to its localized estimation error metric that mirrors the centralized scheme while incurring minimal overhead. Both schemes show stable performance even under adverse conditions such as limited pilot symbols or dense AP-UE associations, and importantly, they enhance user fairness by maintaining strong per-user SE performance, particularly benefiting weak users. Overall, the results across all simulations confirm that the proposed schemes are better performing, scalable, and more robust than various existing schemes.

\section{Conclusion}
\label{conclusion}

This paper introduces two scalable, low-complexity, and dynamic PA schemes for large-scale D-mMIMO systems: EEM-PA  and DPB-PA. The proposed EEM-PA scheme leverages a novel global estimation error metric to guide centralized pilot allocation, achieving superior spectral efficiency across diverse network scenarios. On the other hand, the DPB-PA scheme operates in a fully decentralized manner, requiring only local information without any coordination among access points. Despite its distributed nature, DPB-PA demonstrates strong and competitive performance.

The dynamic and sequential design of both schemes ensures adaptability to varying UE densities, pilot lengths, and AP-UE associations, making them well-suited for practical and large-scale deployments. The EEM-PA scheme is ideal for scenarios with centralized control and sufficient fronthaul capacity. In contrast, the DPB-PA scheme is particularly effective in large deployment where centralized processing is not feasible, offering scalability with minimal signaling overhead. By addressing both centralized and fully distributed architectures, the proposed schemes provide robust, efficient, and flexible solutions to meet the demands of future dynamic and large-scale distributed massive MIMO networks.

\end{document}